\documentclass{czjphys}         
\usepackage{amssymb}
\usepackage{graphicx}
\usepackage{subfigure}
\begin{document}
\title{Impact of Uncertainties in Hadron Production on Air-Shower Predictions}  
%
\authori{T. Pierog, R. Engel and D. Heck }      \addressi{Forschungszentrum Karlsruhe, Institut f\"ur Kernphysik, 76021 Karlsruhe, Germany }
\authorii{}     \addressii{}
\authoriii{}    \addressiii{}
\authoriv{}     \addressiv{}
\authorv{}      \addressv{}
\authorvi{}     \addressvi{}
%
\headauthor{T. Pierog et al.}            
\headtitle{Impact of Uncertainties in Hadron Production on Air-Shower Predictions\ldots}             
\lastevenhead{T. Pierog et al.: Impact of Uncertainties in Hadron Production on Air Shower Predictions\ldots} 
\refnum{A}
\daterec{XXX}    
\issuenumber{?}  \year{2006}
\setcounter{page}{1}
\maketitle

\begin{abstract}
At high energy, cosmic rays can only be studied by measuring the extensive 
air showers they produce in the atmosphere of
the Earth. Although the main features of air showers can be understood within a 
simple model of successive interactions, detailed simulations and a realistic description of particle production
are needed to calculate observables relevant to air
shower experiments. Currently hadronic interaction models are the main source of 
uncertainty of such simulations. We will study the effect of using different hadronic
models available in {\sc corsika} and {\sc conex} on extensive air shower
predictions.
\end{abstract}

\section{Introduction}

Due to the steeply falling energy spectrum of cosmic rays, direct detection by satellite- or balloon-borne instruments is only possible up to about $\sim 10^{14}$ eV. Fortunately, at such high energy, the cascades of secondary particles produced by cosmic rays reach the ground and can be detected in coincidence experiments. The cascades are called extensive air showers and are routinely used to make indirect measurements of high energy cosmic rays. The upper limit of the detectable energy is given by the area and exposure time of the detector. For instance, the Pierre Auger Observatory \cite{Auger}, whose Southern detector is currently under construction in Argentina, is designed to detect particles of $\sim 10^{20}$ eV for which the flux is less than one particle per km$^2$ and century.

Air showers can be observed using different detection techniques. 
The most frequently employed technique is the measurement of secondary particles reaching ground. Using an array of particle detectors (for example, sensitive to $e^\pm$ and $\mu^\pm$), the arrival direction and information on mass and energy of the primary cosmic ray can be reconstructed. The main observables are the number and the lateral and temporal distributions of the different secondary particles. 
At energies above $\sim 10^{17}$ eV, the longitudinal profile of a shower can be directly observed by measuring the fluorescence light induced by the charged particles traversing the atmosphere. Two main observables can be extracted from the longitudinal shower profile: the energy deposit or the number of particles, $N_{\rm max}$, at the shower maximum and X$_{\rm max}$, the atmospheric depth of the maximum. Again, these quantities can be used to estimate the energy and mass of the primary particles. Shower-to-shower fluctuations of all observables provide also very useful composition information.

As a consequence of the indirect character of the measurement, detailed simulations of air showers are needed to extract information on the primary particle from shower observables. Whereas electromagnetic interactions are well understood within perturbative QED, hadronic multiparticle production cannot be calculated within QCD from first principles. Differences in modelling hadronic interactions, which cannot be resolved by current accelerator data, are the main source of uncertainty of air shower predictions \cite{Knapp96a,Knapp:2002vs}. 

In this article, we will discuss the relation between hadronic multiparticle production and air shower observables.
In Sec.~\ref{heitler}, a very simple toy model, based on the model by Heitler \cite{heitler} from the fifties, is used to introduce the basic features of air showers and their relation to multiparticle production. Using detailed Monte Carlo simulations done with {\sc corsika} \cite{corsika} and {\sc conex} \cite{conex} the uncertainty of cosmic ray measurements due to hadronic interaction models is illustrated in Sec.~\ref{results}. Here we concentrate on the uncertainties implied by different models. In addition, each model has a number of parameters that influence the extrapolation to high energy. Studies of the uncertainty range due to variation of model parameters within a single model can be found in \cite{Engel03a,Ostapchenko:2003sj,Zha:2003b,OstapchenkoC2CR}. In general, the differences between models are comparable or bigger than the differences of the predictions that can be obtained within a single model by parameter adjustment. 

\section{Heitler's Model} \label{heitler}         

To qualitatively describe the dependence of shower development on some basic parameters of particle interaction, decay and production, a very simple toy model can be used. Although initially developed for electromagnetic (EM) showers \cite{heitler} it can also be applied to hadronic showers \cite{matt}. 

\subsection{Electromagnetic showers}      

\begin{figure}[hpbt] 
\centerline{
	\subfigure[]{
	\includegraphics[width=0.35\columnwidth,bb=0 0 347 371]{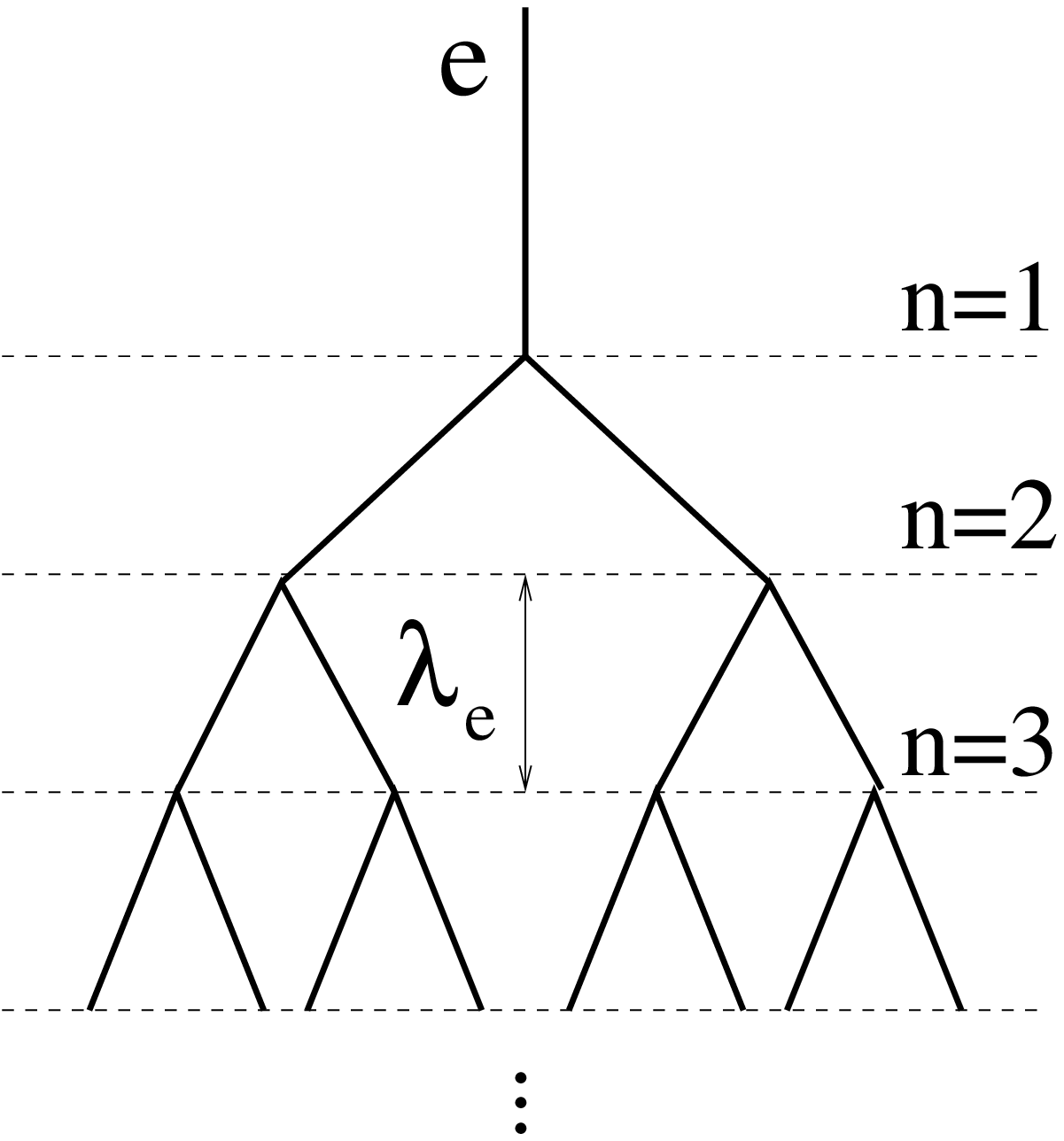}}
	\subfigure[]{
	\includegraphics[width=0.35\columnwidth,bb=0 0 350 385]{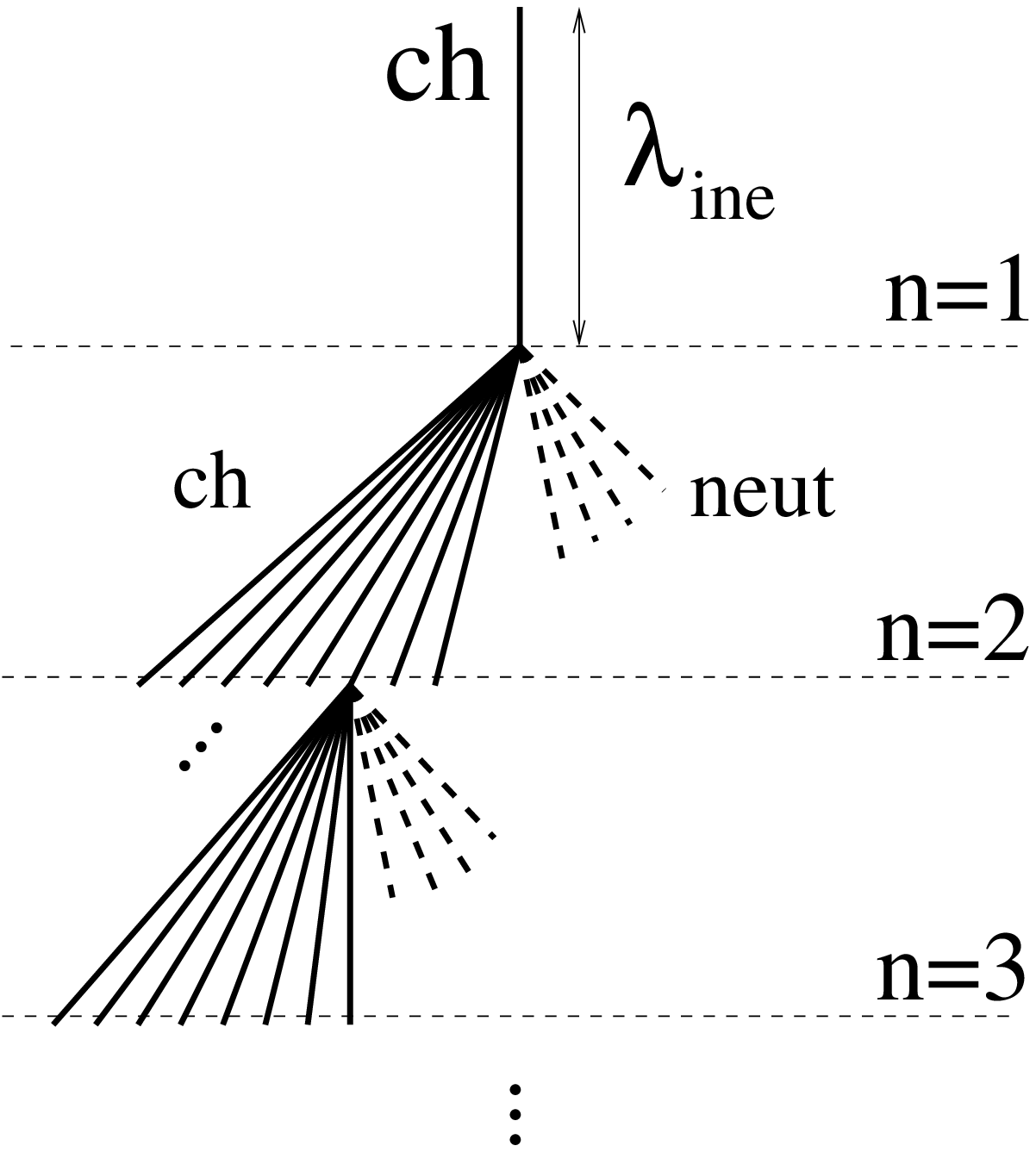}}
}

\caption{\label{diagr}Schematic view of (a) electromagnetic and (b) hadronic cascades. In the latter, dashed lines represent neutral particles ($\pi^0$) and solid lines charged particles ($\pi^\pm$). Only one charged hadron interaction is shown for each generation.}
\end{figure}

For simplicity, instead of having three particle types ($\gamma$, $e^+$ and $e^-$) like in electromagnetic (EM) showers, we will consider only one particle with energy $E$ with only one EM interaction producing two new particles with energy $E/2$ after a fixed interaction length of $\lambda_e$, see Fig. \ref{diagr}(a).

Denoting with $n$ the number of generations (consecutive interactions), the number of particles at a given depth $X=n\cdot\lambda_e$ follows from
\begin{equation}
N(X)=2^n=2^{X/\lambda_e},
\end{equation} 
with the energy $E$ per particle for a given primary energy $E_0$ being
\begin{equation}
E(X)=\frac{E_0}{2^{X/\lambda_e}}.
\end{equation} 

Defining the critical energy $E_c$ ($\sim 85$ MeV in air) as the energy below which energy loss processes dominate over particle production, one can make the assumption that the shower maximum is reached at a depth at which the energy of the secondary particles reaches $E_c$. Then two main shower observables are given by
\begin{equation}
N_{\rm max}=\frac{E_0}{E_c} \quad{\rm and }\quad X^e_{\rm max}(E_0)\sim\lambda_e\cdot\ln\left( \frac{E_0}{E_c}\right). 
\end{equation} 
This simplified picture does not reproduce the detailed behavior of an EM shower, but two important features are well described: the number of particles at shower maximum is proportional to $E_0$ and the depth of shower maximum depends logarithmically on the primary energy $E_0$.

\subsection{Hadronic showers}

Generalizing this idea, a hadronic interaction of a particle with
energy $E$ is assumed to produce $n_{\rm tot}$ new particles with energy $E/n_{\rm
tot}$, two third of which being charged particles $n_{\rm ch}$ (charged pions)
and one third being neutral particles $n_{\rm neut}$ (neutral pions), as shown
Fig.~\ref{diagr}(b).  Neutral particles decay immediately into EM particles
($\pi^0 \rightarrow 2 \gamma$).  After having travelled a distance corresponding to the mean interaction length
$\lambda_{\rm ine}$, charged particles re-interact with air nuclei if their energy
is greater than some typical decay energy $E_{\rm dec}$.
\\
\subsubsection*{{\it Energy transfer}}

In each hadronic interaction, one third of the energy is transferred to the EM
shower component. After $n$ generations the energy in the hadronic and EM
components is given by 
\begin{eqnarray} 
E_{\rm had} & = & \left(\frac{2}{3}\right) ^n  E_0\\
E_{\rm EM} & = & \left[ 1 - \left( \frac{2}{3}\right) ^n\right]  E_0, \label{transf}
\end{eqnarray}
where $n$ will be calculated later. Simulations show that the number of generations is typically about 5 to 6 \cite{MeurereC2CR}.

Even in an air shower initiated by a hadron, most of the energy is carried by EM particles ($\sim90\%$ for $n=6$). Hence the depth of shower maximum is given by that of the EM shower component, $X^e_{\rm max}$. As the first hadronic interaction produces EM particles of energy $\sim E_0/n_{\rm tot}$ one gets
\begin{eqnarray}
X_{\rm max}(E_0) & \sim & \lambda_{\rm ine}+X^e_{\rm max}(E_0/n_{\rm tot}) \\
            & \sim & \lambda_{\rm ine}+\lambda_e\cdot\ln\left( \frac{E_0}{n_{\rm tot}E_c}\right),  \label{eqxmax}
\end{eqnarray}
where $\lambda_{\rm ine}$ is the hadronic interaction length. This simplified expression for the depth of maximum neglects the EM sub-showers initiated by hadrons of later generations. The inclusion of higher hadronic generations does not change the structure of Eq.~(\ref{eqxmax}), only the coefficients change (see, for example, \cite{Alvarez-Muniz:2002ne}).

\subsubsection*{{\it Muon component}}

To keep the picture simple, we assume that all charged hadrons decay into muons when their energy reaches $E_{\rm dec}$. In a real shower, this limit can be seen as the characteristic energy where interaction length and decay length of charged pions are similar (about 150 GeV for pions). By construction, charged particles will reach the energy $E_{\rm dec}$ after $n$ interactions
\begin{equation}
E_{\rm dec}=\frac{E_0}{(n_{\rm tot})^n}.
\end{equation} 
Since one muon is produced in the decay of each charged particle, we get
\begin{equation}
N_{\mu}=n_{\rm ch}^n=\left( \frac{E_0}{E_{\rm dec}} \right) ^ \alpha,
\end{equation} 
with $\alpha=\ln{n_{\rm ch}}/\ln{n_{\rm tot}}\approx0.82\ldots0.95$ \cite{Alvarez-Muniz:2002ne}. The number of muons produced in an air shower depends not only on the primary energy and air density, but also on the charged and total particle multiplicities of hadronic interactions.

It should be kept in mind that the parameters of the model are only
effective quantities and are not identical to the respective
quantities measured at accelerators. In particular, the approximation
of all secondary particles carrying the same energy is only motivated
by the fact that it allows us to obtain simple, closed
expressions. The well-known leading particle effect, typically
quantified by the (in)\-elasticity of an interaction, can be
implemented in the model \cite{matt} but will not be considered here.

\subsubsection*{{\it Superposition model}}

In case of a nucleus being the primary particle, one can use the superposition model to deduce the main observables from the above-written formulas. In this model, a nucleus with mass $A$ and energy $E_0$ is considered as $A$ independent nucleons with energy $E_{\rm h}=E_0/A$. This leads to
\begin{eqnarray}
N^A_{\rm max} & \approx & A\cdot\frac{E_{\rm h}}{E_c}=\frac{E_0}{E_c}=N_{\rm max} \label{nmaxa}\\
X^A_{\rm max} & \approx & X_{\rm max}(E_0/A) \\
N^A_{\mu} & \approx & A\cdot \left( \frac{E_0/A}{E_{\rm dec}} \right) ^\alpha=A^{1-\alpha}\cdot N_{\mu}\label{nmua}.
\end{eqnarray}
Note that, whereas there is no mass dependence of the number of charged particles at shower maximum, both the number of muons and the depth of maximum depend on the mass of the primary particle. The heavier the shower-initiating particle the more muons are expected for a given primary energy.

\section{Uncertainties due to Hadronic Interaction Models} \label{results}

It is clear that the model described above is only giving a very much over-simplified account of air shower physics. However, the model allows us to qualitatively understand the dependence of many air shower observables on the characteristics of hadronic particle production. The parameters of hadron production being most important for air shower development are the cross section (or mean free path $\lambda_{\rm ine}$), the multiplicity of secondary particles of high energy, $n_{\rm tot}$, and the production ratio of neutral to charged particles. Unfortunately, these parameters are not well constrained by particle production measurements at accelerators. Depending on the assumptions on how to extrapolate existing accelerator data, the predictions of hadronic interaction models differ considerably.

\subsection{Hadronic interaction models}

There are several hadronic interaction models commonly used to
simulate air showers. For high energy interactions ($E_{\rm
lab}\gtrsim 100$~GeV), these models are {\sc dpmjet II.55} and III
\cite{dpmjet}, {\sc neXus} 2 and 3.97 \cite{nexus,nexus3}, {\sc qgsjet~01} and II
\cite{qgsjet,qgsjet2,qgsjet22}, and {\sc sibyll~2.1}
\cite{Fletcher94,JEngel92,sibyll}. The physics models and assumptions are discussed
in, for example, \cite{OstapchenkoC2CR}. All the high-energy
interaction models reproduce accelerator data reasonably well but
predict different extrapolations above $E_{\rm cms}\sim$1.8 TeV
($E_{\rm lab}\sim10^{15}$~eV) that lead to very different results at
high energy \cite{Knapp:2002vs,hadronic}. The situation is different at low
energy where several measurements from fixed target experiments are
available \cite{MeurereC2CR}. There one of the main problems is the
extrapolation of measurements to the very forward phase space region
close to the beam direction and the lack of measurements of
pion-induced interactions.  At low energy, models based on data
parameterization and/or microscopic models such as {\sc fluka} \cite{fluka},
{\sc gheisha} \cite{gheisha}, or {\sc u}r{\sc qmd} 1.3 \cite{urqmd} are
used. Differences in the results of those models mainly play a role
for the number of muons far from the shower axis \cite{Drescher:2003gh,low}.

\begin{figure}[hpbt] 
	\begin{center}
	\includegraphics[width=0.9\columnwidth,bb=0 4 554 342]{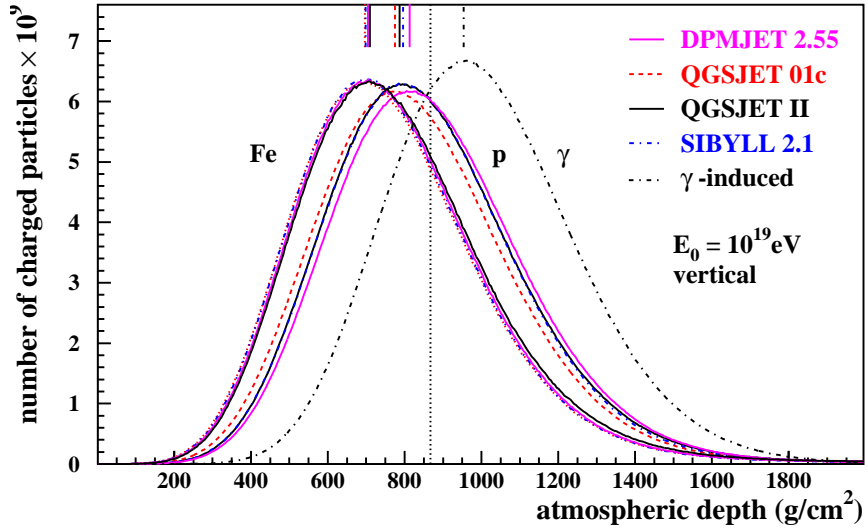}
	\end{center}

\caption{\label{longi}Mean longitudinal profile of vertical iron, proton and $\gamma$ induced showers at $10^{19}$~eV using different high-energy hadronic interaction models (100 showers averaged).}
\end{figure}

In the following we will only focus on high energy interaction models.
A first illustration of differences of air shower predictions due to model uncertainties is shown in Fig.~\ref{longi}. The shower maximum is shifted from one model to another by up to 40 g/cm$^2$. The difference is bigger for proton induced showers than that of iron. The small differences of the predictions of iron showers can partially be explained by the superposition model. Iron showers can be considered as many proton showers of 1/56 of the primary energy since. At lower energy, closer to existing collider data, the model predictions are more similar. On the other hand the number of particles at shower maximum is almost model-independent.

\subsection{Simulation results}

Using the air shower simulation packages {\sc corsika} and {\sc conex}
and various high-energy hadronic interaction models, we can test the
model dependence of different energy and mass estimators of air shower
experiments and get an estimate of the resulting uncertainties.

\subsubsection*{{\it Energy estimation}}

\begin{figure}[hpbt]
	\begin{center}
	\includegraphics[width=0.9\columnwidth,bb= 0 0 567 405]{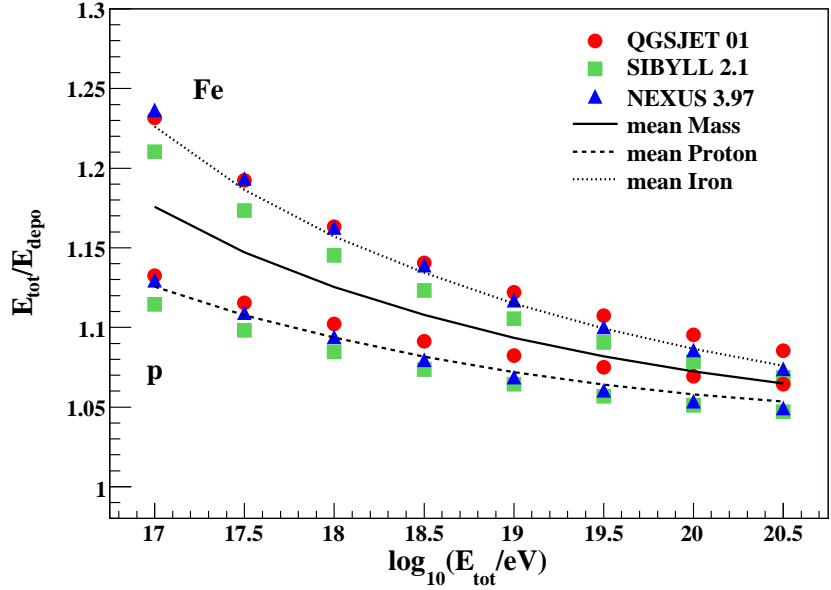}
	\end{center}

\caption{\label{factor}Mean factor for the conversion of observed
(calorimetric) energy to total energy for iron (dotted) or proton
(dashed) induced showers \cite{PierogICRC05}. The conversion factor is
shown for {\sc qgsjet~01} (circles), {\sc sibyll~2.1} (squares) or {\sc neXus}
3.97 (triangles). The mean conversion factor is calculated by
averaging all proton and iron predictions.}
\end{figure}

In case of a fluorescence light detector, the energy is measured by
integrating the signal of the observed longitudinal ionization energy
deposit profile. To account for energy carried by neutrinos and
partially also muons, one has to convert the observed, calorimetric
energy to the total shower energy. The conversion factor is energy-
and mass-dependent. Predictions for the conversion factor obtained
from simulations done with {\sc conex} \cite{PierogICRC05} are shown
in Fig.~\ref{factor}.
From Eq.~(\ref{transf}), we know that this conversion factor depends mainly on the effective number of generations, which in turn is only logarithmically dependent on $n_{\rm tot}$. Therefore, for a given mass, the uncertainty due to the hadronic model is only about 2\%. Furthermore, the unknown mass of the primary particles leads to an uncertainty of less than 3\% on the energy estimation above $10^{19}$~eV.

\begin{figure}[hpbt]
	\begin{center}
	\includegraphics[width=0.9\columnwidth,bb=0 0 567 357]{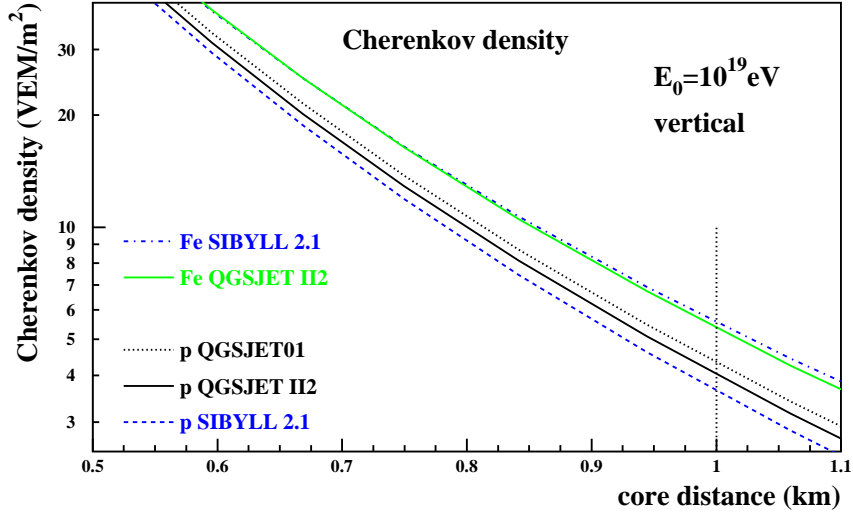}
	\end{center}

\caption{\label{lateral}Mean lateral distribution function of
Cherenkov density for $10^{19}$~eV vertical proton and iron induced
showers and different high-energy hadronic interaction models.}
\end{figure}

Ground based detectors use the density of particles at a given
distance from the shower core to reconstruct the shower energy. In
case of the Pierre Auger Observatory, the Cherenkov density in units
of that of vertical muons (vertical equivalent muon, 1~VEM~$\approx
240$~MeV energy deposit) is measured at a lateral distance of
1000~m. {\sc Corsika} simulations of this quantity are shown for
vertical proton and iron induced showers at 10$^{19}$eV in
Fig.~\ref{lateral}.
In this case, the uncertainty between models for proton induced
showers is about $\pm$7\%. If one takes into account the mass
uncertainty, the theoretical uncertainty increases to about
$\pm$17\%. At least qualitatively, the difference between {\sc qgsjet} and
{\sc sibyll} can be understood in terms of the $\alpha$ parameter of the
Heitler-type model. The $\alpha$ parameter is lower for the {\sc sibyll}
model than for {\sc qgsjet}, 0.84 and 0.85, respectively
\cite{Alvarez-Muniz:2002ne}. This small difference is sufficient to
give a 15\% difference in the number of muons which contribute about
50\% to the Cherenkov density.

\subsubsection*{{\it Mass estimation}}

It is desirable to estimate both the energy and the mass of the
primary particles at the same time. First of all the composition of
cosmic rays holds important clues on the nature of their sources
propagation processes. Furthermore one can improve the energy
reconstruction considerably by knowing the mass of the primary
particle.

\begin{figure}[hpbt]
	\begin{center}
	\includegraphics[width=0.9\columnwidth,bb= 0 0 567 383]{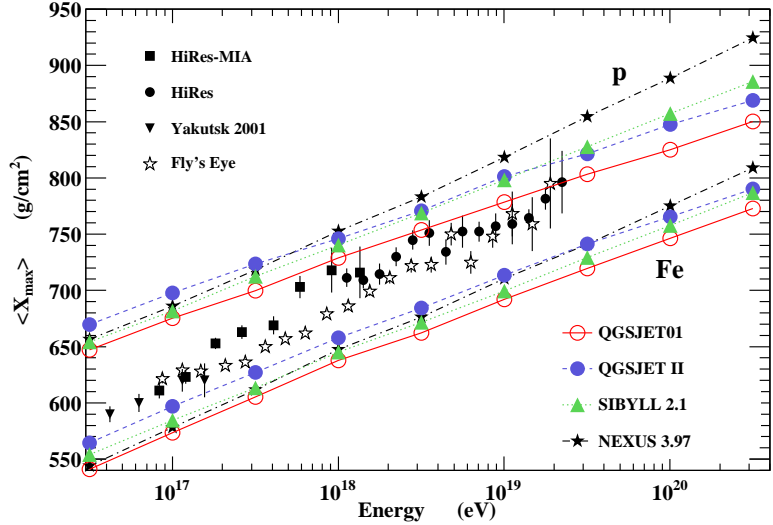}
	\end{center}

\caption{\label{xmax}Mean X$_{\rm max}$ for proton and iron induced showers as a function of the primary energy. Predictions of different high-energy hadronic interaction models are compared to data. Refs. to the data can be found in \cite{Engel:2004ui}.}
\end{figure}

In an experiment employing the fluorescence light technique, the mass
of the primary particle can be estimated through the measurement of
the depth of shower maximum. In Fig.~\ref{xmax}, {\sc conex}
simulations of the mean depth of shower maximum, $\langle X_{\rm
max}\rangle$, are shown together with a compilation of measurements.
The model dependence of $\langle X_{\rm max}\rangle$ is obvious. The
mean depth of shower maximum  mainly depends on the hadronic inelastic
cross section, the multiplicity as well as the inelasticity, see
Eq.~(\ref{eqxmax}). For example, {\sc neXus} (dash-dotted line with
stars) is characterized by the highest inelastic cross section and the
lowest multiplicity. Therefore it is not surprising that it predicts
the highest $\langle X_{\rm max}\rangle$. On the other hand, {\sc qgsjet~01}
(full line with open circles) has the lowest inelastic cross section
and the highest multiplicity, which explains why its $\langle X_{\rm
max}\rangle$ is small. The difference between the model predictions is
limiting the conclusions that can be drawn from the measurements. For
example, in a simple interpretation of the experimental data at the
highest energy, using {\sc qgsjet~01} will lead to a composition dominated
by protons rather than a mixed composition as would be obtained if
{\sc neXus} is taken as reference.

\begin{figure}
	\begin{center}
	\includegraphics[width=0.9\columnwidth,bb= 0 0 567 430]{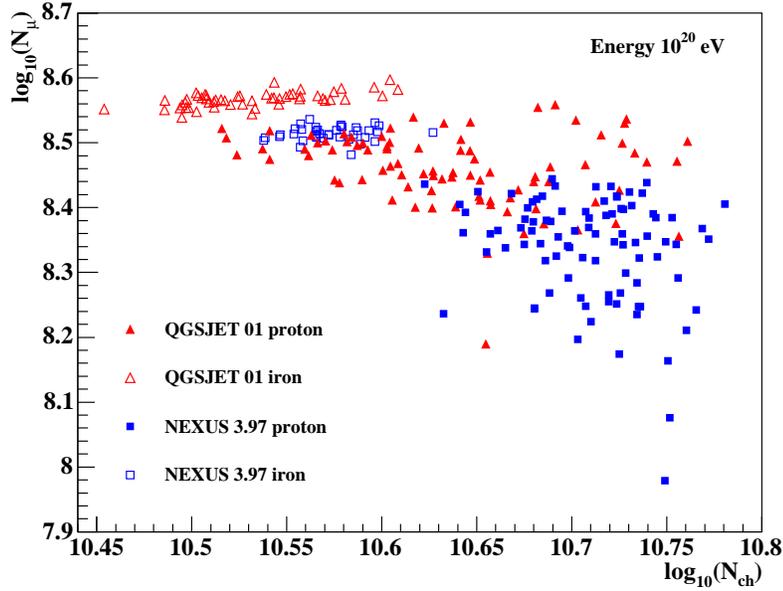}
	\end{center}

\caption{\label{contour} Correlation of charged particle and muon multiplicites of proton (filled symbols) and iron (open symbols) induced showers at $10^{20}$~eV. Each point represents an individual shower simulated with {\sc conex}. Triangles show results obtained with {\sc qgsjet~01} and squares that of {\sc neXus}~3.97.}
\end{figure}

The mass dependence of the correlation of the charged particle and
muon multiplicities of showers is used in surface array experiments
for both energy and composition reconstruction.  According to the
prediction of the superposition model, see Eqs.~(\ref{nmaxa}) and
(\ref{nmua}), one would expect a separation of proton and iron induced
showers at fixed energy in the $N_{ch}-N_{\mu}$ plane. However, the
sensitivity of this method of mass reconstruction is limited by its
model dependence and shower-to-shower fluctuations. In
Fig.~\ref{contour}, we show the charged particle and muon shower sizes
of 50 proton and iron showers at Auger observation level (1450m
a.s.l.). Each point corresponds to a 30$^{\circ}$ inclined shower
simulated at $10^{20}$~eV. For clarity, the predictions of only two
models are shown, {\sc qgsjet~01} (triangles) and {\sc neXus}
(squares). There is a clear separation between proton (full symbols)
and iron (open symbols) initiated showers for each model. The number
of electrons at ground is slightly higher for proton induced showers
because $X_{\rm max}$ is larger and the maximum is closer to the
ground. As expected, the number of muons in iron induced showers is
higher than in proton ones by about a factor of 1.4 which corresponds
to $A^{1-\alpha}$ with $\alpha \sim 0.92$.


Applying another model, for instance {\sc neXus}~3.97 ({\sc sibyll} gives
similar results in this case), we notice that proton and iron
initiated showers are even more separated than in {\sc qgsjet} simulations
($\alpha \sim 0.86$ is lower and the difference between $\langle
X_{\rm max}\rangle$ for proton and iron is larger). However, iron
induced showers from {\sc neXus} populate a region in the plot that
overlaps that of proton induced showers from {\sc qgsjet~01}. This model
dependence makes it very difficult to interpret experimental
data. Using not only the mean multiplicities but also the
shower-to-shower fluctuations and correlations in the analysis, one
can try to estimate the mass within a given hadronic interaction model
\cite{ulrich}, and even exclude models if the data can not be
reproduced at all. Finally it should be noted that the fluctuations
and also the model dependence are somewhat reduced if particle
densities at about 600 to 1000m from the shower core are considered
instead of total multiplicity (see, for example,
\cite{Hillas,OstapchenkoC2CR}).

Finally it should be mentioned that there are a number of alternative high energy extrapolations of hadronic multiparticle production suggested in literature (for example, see \cite{Pajares,Drescher:2004sd,DiasDeDeus}) that have not been discussed in this work. It can be expected that the uncertainty of the composition reconstruction increases considerably if these scenarios are included. On the other hand, the energy reconstruction by means of fluorescence light detection is very robust and will change only slightly. 

\section{Summary}

Using a simple cascade model, it is possible to find the main
parameters of hadronic interactions that influence air shower
predictions. These parameters, name-ly the inelastic cross sections,
the secondary particle multiplicity, the inelasticity, and the ratio
of charged to neutral hadrons, depend of the hadronic interaction
model. As a consequence, realistic simulations of hadron induced
air-showers are model-dependent, leading to theoretical uncertainties
in the analysis of experimental data. For a ground based detector, the
model-related systematic error on energy estimation can be as large as
17\% at $10^{19}$~eV if the mass of the primary particle is
unknown. The theoretical uncertainties of the energy reconstruction
are much smaller for fluorescence light detectors (less than 5\% even
for unknown primary particle mass). The model dependence of the
primary mass estimation is crucial and currently the mass composition
can only be derived for a given hadronic model. On the other hand,
cosmic ray data of multi-observable detectors can be used to test
hadronic interaction  models at energies much higher than those
reached in laboratory.

\noindent
{\bf Acknowledgements} The authors thank Sergey Ostapchenko and Michael Unger for fruitful discussions on subjects related to this work.

\end {document}